\documentclass[twocolumn,showpacs,preprintnumbers,prl,superscriptaddress,amsmath,amssymb]{revtex4}
\usepackage{graphicx}% Include figure files
\usepackage{dcolumn}% Align table columns on decimal point
\usepackage{bm}% bold math

\begin{document}

%\preprint{}

\title{Control of the direction and rate of nuclear spin flips in InAs quantum dots using detuned optical pulse trains}

\author{S. G. Carter}
\affiliation{Naval Research Laboratory, Washington, DC 20375-5322, USA}
\author{A. Shabaev}
\affiliation{School of Computational Sciences, George Mason University, Fairfax, VA 22030, USA}
\author{Sophia E. Economou}
\affiliation{Naval Research Laboratory, Washington, DC 20375-5322, USA}
\author{T. A. Kennedy}
\affiliation{Naval Research Laboratory, Washington, DC 20375-5322, USA}
\author{A. S. Bracker}
\affiliation{Naval Research Laboratory, Washington, DC 20375-5322, USA}
\author{T. L. Reinecke}
\affiliation{Naval Research Laboratory, Washington, DC 20375-5322, USA}

\date{\today}

\begin{abstract}
We find that detuning an optical pulse train from electronic transitions in quantum dots controls the direction of nuclear spin flips. The optical pulse train generates electron spins that precess about an applied magnetic field, with a spin component parallel to the field only for detuned pulses. This component leads to asymmetry in the nuclear spin flips, providing a way to produce a stable and precise value of the nuclear spin polarization. This effect is observed using two-color, time-resolved Faraday rotation and ellipticity.\end{abstract}

\pacs{78.67.Hc, 78.47.jc, 72.25.Fe }

%78.67.Hc Optical properties of low-dimensional, mesoscopic, and nanoscale materials and structures: Quantum dots
%78.47.jc Time resolved spectroscopy (> 1 psec)   
%72.25.Fe Optical creation of spin polarized carriers   

\maketitle

Spins in semiconductor quantum dots (QDs) are promising quantum bits, with spin coherence times of a few microseconds \cite{Petta_Science05,Greilich_Science06}, controlled interactions with other QDs \cite{Stinaff_Science06}, and fast optical initialization and control \cite{Atature_Science06, Xu_PRL07, Greilich_PRL06, Berezovsky_Science08}. A point of considerable current interest for QDs as qubits is the hyperfine interaction with the many ($10^4$-$10^5$) nuclei in the dot \cite{Merkulov_PRB02, Khaetskii_PRL02, Coish_PRB04, Bracker_PRL05, Stepanenko_PRL06, Deng_PRB06, Witzel_PRB06, Yao_PRB06, Korenev_PRL07, Greilich_Science07, Makhonin_APL08, Reilly_Science08, Danon_PRL08}. The nuclear spin configuration is typically random and varying in time. For a single QD measured over seconds, the nuclei sample many configurations, changing the electron spin splitting through the Overhauser effect. This leads to an apparent electron spin dephasing time $T_2^*$ of a few nanoseconds and an unknown spin splitting.

To overcome this problem, researchers have examined pumping nuclear spins into a narrow distribution of states. In electrically defined quantum dots, repetitive gate manipulation has led to narrower nuclear spin distributions and lengthening of $T_2^*$ \cite{Reilly_Science08}. Several ideas for optically preparing nuclei have been suggested \cite{Stepanenko_PRL06, Korenev_PRL07}, with a number of experiments demonstrating some control over the nuclear spin polarization \cite{Bracker_PRL05, Greilich_Science07, Makhonin_APL08}. In particular in Ref.~\cite{Greilich_Science07}, periodic excitation by an optical pulse train was used in an ensemble of QDs to vary the nuclear spin flip rate as a function of the electron spin precession frequency. The underlying physics there is that sudden changes in the electron spin state due to optical pulses induce nuclear spin flips. Electron spins synchronized to a multiple of the laser repetition rate are not affected by the laser pulses and have much lower nuclear spin flip rates than those not synchronized. With no electron spin polarization along the magnetic field direction, the nuclei flip up or down with equal probability. Thus, the nuclear polarization takes a random walk, as does the Overhauser shifted electron spin precession frequency, until reaching a synchronized precession frequency with a lower nuclear spin flip rate.

In this Letter, we show that energy detuning of the optical pumping can be used to control the sign and rate of nuclear spin flips, leading to more deterministic nuclear spin dynamics. This optical control takes advantage of recent progress in spin manipulation using both real excitation of electronic transitions \cite{Shabaev_PRB03, Economou_PRB05, Dutt_PRL05, Greilich_PRL06, Wu_PRL2007} and virtual excitation that rotates existing spin polarizations \cite{Pryor_APL06, Economou_PRL07, Berezovsky_Science08, Carter_PRB07, KMF_nphys08}. Using two-color, time-resolved Faraday rotation and ellipticity, we examine an ensemble of QD spins at varying detunings from a pump pulse train. The pulse train generates a spin polarization that precesses about an applied magnetic field and, for detuned QDs, also rotates the spins to give a significant component parallel to the magnetic field. This parallel component gives asymmetry in the spin flip rates. For negative detuning, the asymmetry pushes electron spins toward synchronized frequencies. For positive detuning, this component pushes electron spins away from synchronized frequencies, resulting in asymmetry in the spin amplitude versus QD detuning. These results open the way for precise control of the nuclear polarization and a better understanding of the role of the nuclei in electron spin manipulation.

\begin{figure}
\includegraphics{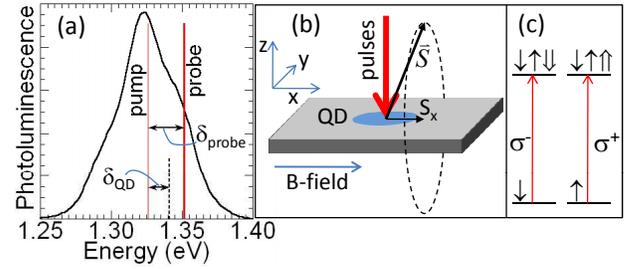}
\caption{\label{fig1} (color online) (a) Photoluminescence of the quantum dots. The red vertical lines represent the pump and probe, and the dashed vertical line represents an arbitrary QD energy. The actual detunings are much smaller than those displayed. (b) Experimental geometry showing spin precession. (c) Electron-trion level diagram, showing the two electron and trion spin states and the allowed transitions. Single (double) arrows are electron (hole) spins.}
\end{figure}

The experiments are performed on a sample consisting of 20 layers of InAs QDs, grown by Molecular Beam Epitaxy through Stransky-Krastanov self-assembly. The QDs were grown using the In-flush technique \cite{Wasilewski_JCG99}, giving a truncated pyramid structure of height $2.5$ nm and lateral dimensions varying from 10-20 nm. We find that a significant fraction (roughly 50\%) of the QDs are singly charged with electrons coming from impurities. The photoluminescence (PL) of the QD sample at $\sim$$5$ K is given in Fig.~\ref{fig1}(a), showing a broad spectrum due to inhomogeneity, with a full-width at half-maximum (FWHM) of $\sim$50 meV.

Electron spin dynamics are measured with two-color time-resolved Faraday rotation (TRFR) and ellipticity (TRFE). A circularly polarized pump laser spectrally fixed at the center of the PL (see Fig.~\ref{fig1}(a)) excites a distribution of QDs with varying detunings $\delta_{QD}$ from the pump. A delayed, vertically polarized probe laser with variable detuning from the pump $\delta_{probe}$ measures the pump-induced polarization rotation or ellipticity for a distribution of QDs near the probe photon energy. Rotation (ellipticity) is due to the induced phase (amplitude) difference between the $\sigma^+$ and $\sigma^-$ components of the probe. TRFR and TRFE are both sensitive to the spin polarization along the optical axis but have different spectral dependences.

Both lasers are wavelength tunable mode-locked Titanium:Sapphire lasers operating at a repetition rate of 81~MHz. The pump laser is set to a photon energy of 1.326 meV with a bandwidth of $0.6$ meV, corresponding to a Fourier-limited pulsewidth of 3 ps, and the probe laser has a bandwidth of $1.3$ meV, corresponding to $1.4$ ps. The pump (probe) is focused onto the sample to a diameter of $\sim$80 $\mu$m ($\sim$50 $\mu$m), with a typical average probe intensity of $\sim$20 W$/$cm$^2$.

% The lasers are phase-locked together with a timing jitter of a few picoseconds. The timing between the pump and probe lasers varies as the probe wavelength is changed. The zero delay for each delay scan is then determined by a spike in the TRFR signal when the pump and probe temporally overlap.

%~0.6mW probe before windows -> ~0.45mW after windows before sample. Gives 22.9W/cm^2 -> round to 20

%Figure \ref{fig1}(b) displays a schematic of the TRFR and TRFE experimental setup. A circularly polarized pump generates spin polarized carriers, and a delayed, vertically polarized probe is transmitted through the QDs. The probe then passes through a half-wave plate, rotating the polarization by $45^\circ$; is split by a polarizing beamsplitter; and sent to an amplified photodiode bridge. This configuration measures polarization rotation, and ellipticity is measured by placing a quarter-wave plate in before the half-wave plate.

\begin{figure}
\includegraphics{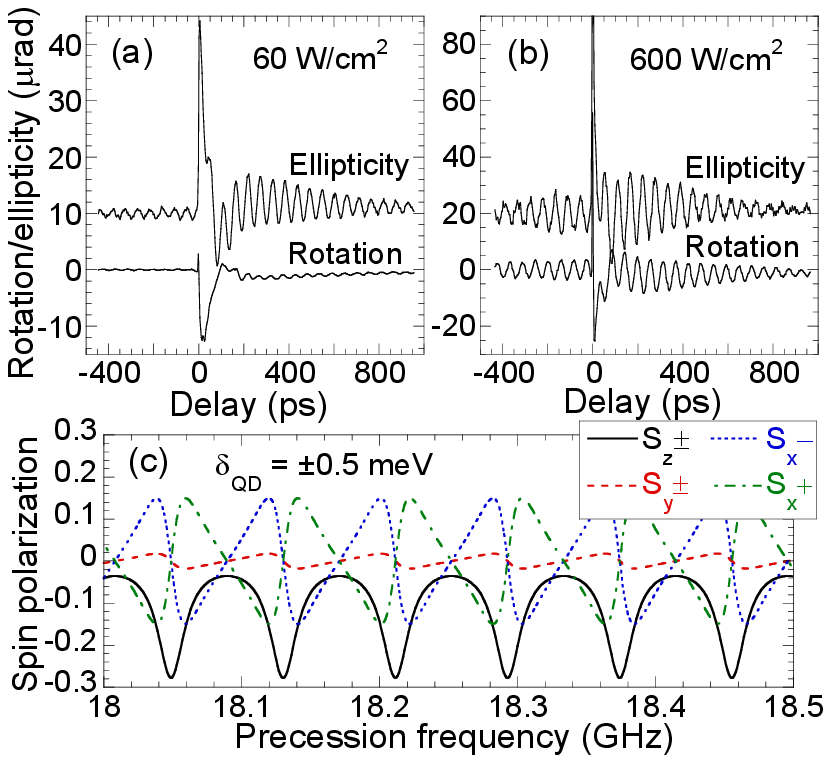}
\caption{\label{fig2} (color online) (a,b) Time-resolved Faraday rotation and ellipticity for zero pump-probe detuning, taken at an average pump intensity of (a) $\sim$60 W$/$cm$^2$ and (b) $\sim$600 W$/$cm$^2$ for B=3 T. The ellipticity curves are offset for clarity. (c) Calculated components of the electron spin polarization (labeled by direction and by the sign of detuning) as a function of the spin precession frequency, for QDs detuned $\delta_{QD}=\pm0.5$ meV from pump pulses of area $\pi$. Complete spin polarization corresponds to $\left|\bm{S}\right| = 0.5$.}
\end{figure}

Figure \ref{fig2}(a) displays TRFR and the TRFE for $\delta_{probe}=0$ at an average pump intensity of $\sim$60 W$/$cm$^2$. The  magnetic field is 3 T, perpendicular to the optical axis (see Fig.~\ref{fig1}(b)). The oscillating signal is the $z$ component of the electron spin, measured by the probe as it precesses about the external magnetic field. The precession frequency is $18.2$ GHz, corresponding to a $\left|g_e\right|$ of $0.43$. The decay time of 450 ps is due to inhomogeneity in $g_e$, confirmed by our magnetic field studies. The TRFR shows similar behavior to the TRFE, but with a weaker signal. TRFR and TRFE measure the real and imaginary part of the susceptibility, i.e., dispersion and absorption respectively.  These are generally known to be odd and even functions of the detuning from the transition. Thus, at $\delta_{probe}=0$, the TRFR should be near zero, and the TRFE should be at a maximum.

The electron spin polarization is generated by optical pumping through the trion state. As shown in Fig.~\ref{fig1}(c), where the spin states are shown in the $z$-basis (along the optical axis), $\sigma^+$ light depletes the $|\uparrow\rangle$ e spin state, leaving behind excess population in the $|\downarrow\rangle$ state. The generated polarization persists after trion recombination if the spin precession period is less than the recombination time \cite{Shabaev_PRB03, Economou_PRB05, Dutt_PRL05}, which is the case in our experiment.

The weaker oscillations observed for negative delays are due to mode-locking of spins \cite{Greilich_Science06}. If the individual electron spin coherence time $T_2$ is longer than the pulse repetition period $T_R$, there is constructive interference for spins that satisfy the phase synchronization condition (PSC) $\omega = 2\pi N/T_R$, where $N$ is an integer and $\omega$ is the spin precession frequency. The negative delay signal is fairly weak for low pump intensities, as in Fig.~\ref{fig2}(a), and stronger at higher pump intensities, as in Fig.~\ref{fig2}(b). By measuring mode-locking for several different values of $T_R$, we estimate $T_2$ at 100-200 ns for this sample, an order of magnitude smaller than the $T_2$ measured in Ref.~\cite{Greilich_Science06}. We assign this difference to the smaller volume of our QDs \cite{Khaetskii_PRL02} compared to those of Ref.~\cite{Greilich_Science06}. 

% and to the differing Indium content \cite{Petrov_PRB08}. 

Surprisingly, the TRFR signal is comparable to the TRFE signal at the higher pump intensity (Fig.~\ref{fig2}(b)), even though the TRFR is expected to be near zero for $\delta_{probe}= 0$. This result is the first hint of physics beyond simply pumping electrons into the $|\downarrow\rangle$ state through real transitions. For $\delta_{QD}\neq 0$, the pulse train also rotates spins about the optical axis through virtual transitions to the trion. As a result, in addition to the spin components perpendicular to the magnetic field, a parallel or antiparallel component also will be generated. Consider for example a spin oriented along $-\hat{z}$ by an initial pulse, which then precesses until the next pulse. If the precession frequency is not at a PSC, there will be an $S_y$ component just before the next pulse, which will be partially rotated into the $\hat{x}$ direction by the virtual process. Figure \ref{fig2}(c) displays the calculated steady state electron spin vector components right after a pulse as functions of the spin precession frequency. The troughs in the $z$-component of the spin vector, $S_z$ correspond to frequencies meeting the PSC. The sign of the detuning determines the sign of the rotation angle \cite{Economou_PRL07}, and thus the sign of $S_x$, without affecting $S_z$ or $S_y$.

\begin{figure} [htp]
\includegraphics{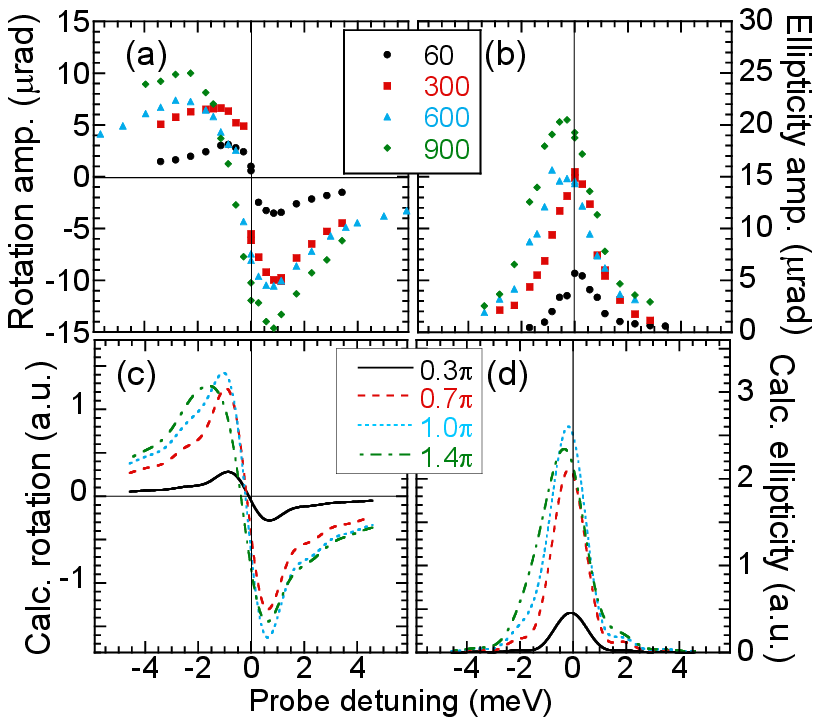}
\caption{\label{fig3} (color online) (a,b) Positive delay amplitude of the experimental (a) TRFR and (b) TRFE as a function of probe detuning from the pump for a series of pump intensities (in W$/$cm$^2$). (c,d) Theoretical calculation of the positive delay amplitude of (c) rotation and (d) ellipticity for a series of pulse areas (shown in the insets).}
\end{figure}

The nuclear dynamics are significantly changed by a nonzero $S_x$ since the nuclear spin flip rates $w_{\pm}$ are proportional to $(1\pm 2 S_x)$ \cite{DyakonovPerel74}. When $S_x \neq 0$ the nuclear spin flip rate from spin up to down, $w_{-}$, is different from the rate to flip from down to up, $w_{+}$ \cite{foot1}. Since $S_x$ depends on $\delta_{QD}$, we expect some manifestation of the changing nuclear dynamics when different energy QDs are probed.

%The dependence of $S_x$ on the detuning implies that these effects can have a manifestation in optical experiments which are sensitive to the spectral position of the lasers. 

%This can be viewed as a pulse-assisted electron-nuclear flip-flop process, present only for QDs which are detuned from the pump.

%In our experiments the probe is swept symmetrically about the pump, which is spectrally fixed. The probe picks up the signal from a range of QDs. By scanning the probe over a wide range of frequencies we measure the relative contribution of the positively and the negatively detuned quantum dots. 

The measured Faraday rotation and ellipticity are plotted versus $\delta_{probe}$ in Fig.~\ref{fig3}(a) and \ref{fig3}(b) respectively, for a series of pump intensities. The rotation and ellipticity are very nearly the expected odd and even functions of $\delta_{probe}$ at the lowest pump intensity, 60 W$/$cm$^2$, indicating the distribution of spin-polarized QDs is symmetric and spectrally narrow.  With increasing pump intensity, the spectral features in Fig.~\ref{fig3}(a,b) are broader and clearly shift toward lower probe energies. The broadening can be explained by higher pump intensities polarizing wider spectral distributions of QDs. The spectral shifts of $\sim$$0.4$ meV for the ellipticity and $\sim$$0.7$ meV for the rotation at the highest pump intensity are more interesting. Certainly these shifts explain the large rotation signal observed at high pump intensities for $\delta_{probe}=0$ in Fig.~\ref{fig2}(b). We attribute the spectral shifts to different nuclear dynamics for positive and negative $\delta_{QD}$.
  
\begin{figure} [htp]
\includegraphics{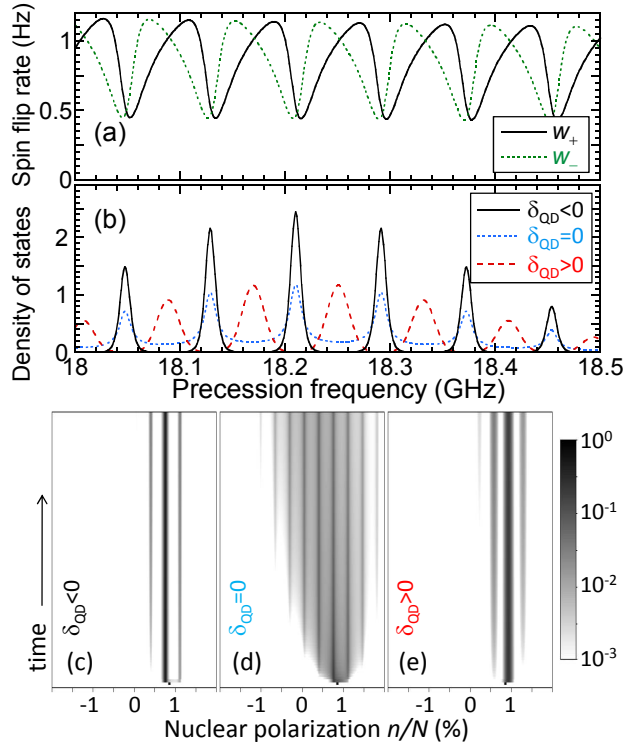}
\caption{\label{fig4} (color online) (a) Calculated nuclear spin flip rates as a function of electron spin precession frequency for $\delta_{QD} = -0.5$ meV and $\pi$-pulses. (b) Calculated steady state density of states as a function of electron spin precession frequency for $\delta_{QD} = -0.8, 0, +0.8$ meV and $\pi$-pulses. For negative $\delta_{QD}$ the QDs are focused toward the PSCs more efficiently than for the $\delta_{QD}=0$ case, while for positive $\delta_{QD}$ the QDs are pushed away from the PSCs. (c,d,e) Time evolution of $P(n)$ for (c) $\delta_{QD} <0$, (d) $\delta_{QD} = 0$, and (e) $\delta_{QD} > 0$, each using $\pi$-pulses. The logarithmic scale for $P(n)$ is cut off for $P(n)<10^{-3}$.}
\end{figure}

Our calculations, which include the feedback effect of the nuclei on the electron spin precession frequency, qualitatively reproduce these shifts, as shown in Fig.~\ref{fig3}(c,d). To calculate the optical response from the inhomogeneous QD ensemble we first find the steady state expressions for the spin components. We use pulse shapes for the pump that provide analytical solutions for any pulse strength and detuning. Our calculations have been done for hyperbolic secant pulses \cite{RosenZener} and for square pulses, and they give qualitatively the same results. The nuclear spin flip rates are functions of the precession frequency $\omega$, the Rabi frequency $\Omega$, the detuning $\delta_{QD}$, and the spin vector components, which are also functions of $\omega$, $\Omega$, and $\delta_{QD}$ \cite{Greilich_Science07}:
\begin{eqnarray}
w_{\pm} \propto  \frac{W(\Omega,\delta_{QD})}{2 T_R}\frac{\left[1+2 S_z \right]}{\omega^2}\left[1\pm 2S_x\right],
\label{rates}
\end{eqnarray}
where $W$ is the transition probability of the allowed electron to trion transition. The rates are plotted in Fig. \ref{fig4}(a) as functions of the precession frequency. Using Eq. (\ref{rates}), we solve numerically for the steady state nuclear polarization probability distribution, $P(n)$, where $n$ is the number of spin up nuclei minus the number spin down. The total number of nuclei $N$ is estimated at $20,000$ from the size of our QDs. The time evolution of $P(n)$ is displayed in Figs. \ref{fig4}(c-e), starting from a very narrow initial distribution centered at $n/N=0.85\%$, for detuned and resonant pulses. From the steady state nuclear polarization distribution we find the density of states (DOS) of the electronic precession frequencies for a series of detunings for each pump pulse area.  Fig. \ref{fig4}(b) shows the results of our calculation; the dramatic effect of the detuning is seen by comparing the DOS of the positive with that of the negative detuning. For negative $\delta_{QD}$ the DOS is concentrated at the PSCs, giving good mode-locking, \textit{i.e.} the net $S_z$ component averaged over all the QDs is large. For positive $\delta_{QD}$ the DOS is concentrated in-between the PSCs, giving poor mode-locking. This asymmetry gives rise to the spectral shift in the calculated rotation and ellipticity spectra in Fig. \ref{fig3}(c,d). 

An intuitive description of the time dynamics that lead to this kind of effect can be given based on Figs. \ref{fig2}(c) and \ref{fig4}(a). Consider a negatively detuned QD with a precession frequency close to a PSC. If $\omega$ is bit smaller (larger) than the PSC, $S_x$ will be positive (negative) and make nuclear spins more likely to flip up (down) \cite{foot}. In both cases, the nuclear polarization will change to move $\omega$ toward the PSC. For a positively detuned QD, the sign of $S_x$ is opposite, so the nuclear polarization changes to move $\omega$ away from the PSC. The positively detuned QDs settle at `anti-synchronized' frequencies, as shown in Fig.~\ref{fig4}(b), which results in a much lower spin polarization. Note the nearly zero density between stable frequencies for detuned QDs of either sign compared to resonant QDs.  

These results have important implications not only for the electronic spin qubit but also for controlling the nuclear spins, narrowing their distribution, and perhaps making possible their use for the storage of quantum information \cite{taylor}. The nuclear dynamics are more controlled and the resulting distribution more stable for detuned pulses compared to resonant pulses, due to the directionality of nuclear spin flips. In Fig.~\ref{fig4}(c), ($\delta_{QD}<0$), the nuclear spin almost immediately jumps to a nearby stable polarization, corresponding to a PSC, with very little leakage or drift into other PSCs compared to the resonant case. By slowly changing the repetition rate of the laser $T_R$ it should be possible to strongly polarize the nuclei. In conclusion, we have shown that by using optically detuned pulses we can control the effects of the hyperfine interaction of the electron with the nuclear spin. This work provides an additional handle toward optical control of the nuclear QD spins.

This work is supported by the US Office of Naval Research. One of us (S.E.E.) is an NRC/NRL Research Associate.
\bibliography{references}

\begin{thebibliography}{34}
\bibitem[{foo({\natexlab{a}})}]{foot1}
\bibinfo{note}{We treat each nucleus as a spin-1/2 system. This provides
  considerable simplification but still captures the main physics.}
\bibitem[{foo({\natexlab{b}})}]{foot}
\bibinfo{note}{We assume clockwise spin precession about the magnetic field.
  Counterclockwise spin precession inverts the sign of $S_x$ and therefore the
  dependence on $\delta_{QD}$. From the direction of the measured spectral
  shift we verify the precession direction and determine the sign of the g
  factor.}
\end{thebibliography}
\end{document}